\documentclass[10pt,a4]{article}

\newcommand{\be}{\begin{equation}}
\newcommand{\ee}{\end{equation}}
\newcommand{\bi}[1]{\vspace{-3mm} \bibitem{#1}}

\begin{document}
\begin{center}
{\Large \it Physics Letters A 288 (2001) 173-182}
\end{center}
\vskip 5 mm

\begin{center}
{\Large \bf Quantization of 
non-Hamiltonian and Dissipative Systems}\\
\vskip 5 mm
{\Large \bf Vasily E. Tarasov } \\
\vskip 5 mm
Theoretical High Energy Physics Department, Skobeltsyn 
Institute of Nuclear Physics, \\
Moscow State University, Moscow 119992, Russia  \\
\vskip 3 mm
E-mail: tarasov@theory.sinp.msu.ru
\end{center}

\vskip 5 mm
\begin{abstract}
{\normalsize 
A generalization of canonical quantization
which maps a dynamical
operator to a dynamical superoperator
is suggested. Weyl quantization of dynamical
operator, which cannot be represented as Poisson bracket
with some function, is considered. The usual
Weyl quantization of observables is a specific case of suggested quantization.
This approach allows to define consistent quantization
procedure for non-Hamiltonian and dissipative systems.
Examples of the harmonic oscillator with friction 
(generalized Lorenz-Rossler-Leipnik-Newton equation), the
Fokker-Planck-type system and Lorenz-type system are considered.
}
\end{abstract}



\twocolumn

\section{Introduction}

The quantization of dissipative and non-Hamiltonian classical
systems is of strong theoretical interest. As a rule,
any microscopic system is always embedded in some
(macroscopic) environment and therefore it is never really
isolated. Frequently, the relevant environment is in principle
unobservable or it is unknown \cite{E1}-\cite{E3}.
This would render theory of dissipative and non-Hamiltonian systems
a fundamental generalization of quantum mechanics \cite{Prig}.

We can divide the most frequent methods of quantization of
dissipative and non-Hamiltonian systems into two groups.
The first method uses a procedure of doubling of
phase-space dimension \cite{B1}-\cite{B3}.
The second method consists in using an explicitly time-dependent
Hamiltonian \cite{K1}-\cite{K7}.

Bateman has shown \cite{B1} that in order to use the usual
canonical quantization methods a procedure of doubling of
phase-space dimension is required. To apply the usual
canonical quantization scheme to dissipative and non-Hamiltonian
systems, one can double the numbers of degrees of
freedom, so as to deal with an effective isolated system.  The
new degrees of freedom may be assumed to represent by
collective degrees of freedom of the bath with absorb the energy
dissipated by the dissipative system \cite{B2,B3}.

Cardirola  \cite{K1} and Kanai \cite{K2} have shown that it may
be possible to put the equation of motion for dissipative system
into time-dependent Hamiltonian form and then quantize them
in the usual way \cite{K1}-\cite{K7}. However, the
corresponding canonical commutation relations violate the
uncertainty principle \cite{Edw}. The reason for this violation
would appear from the explicit
dependence of Hamiltonian and momentum on the time.

To construct a quantization of dissipative and non-Hamiltonian
systems consistently, it is possible to exceed the
limits of Lie algebras and groups. The condition of
self-consistency for a quantization of dissipative systems
requires the application of commutant-Lie (Valya) algebra
\cite{Tartmf4,Tartmf3}. Unfortunately, these algebra and its
representation have not been thoroughly studied.

Note \cite{Dys,K7} that Feynman wanted to develop a
procedure to quantize classical equation of motion without
resort to a Hamiltonian. It is interesting to quantize a classical
system without direct reference to a Hamiltonian.
A general classical system is most easily defined in terms of
its equations of motion.
In general case it is difficult to determine whether a
Hamiltonian exists, whether it is unique if it does exist,
and what its form is if it exists and is unique \cite{DMS,Hen}.
Therefore, quantization that bypasses direct reference to a
Poisson bracket with some Hamiltonian may have practical advantages.


Canonical quantization defines a map
of real functions into self-adjoint operators \cite{Ber1,Ber2}.
A classical observable is described by some real function
$A(q,p)$ from a function space ${\cal M}$.
Quantization of this function leads to self-adjoint
operator $\hat A(\hat q, \hat p)$ from some operator space $\hat {\cal M}$.
Classical state can be described  by non-negative-normed
function $\rho(q,p)$ called density distribution function.
Quantization of a function $\rho(q,p)$ leads to non-negative
self-adjoint operator $\hat \rho$ of trace class
called matrix density operator. This description allows to consider
a state as a special observable.

Time evolution of an observable $A_t(q,p)$ and a state $\rho_t(q,p)$ in
classical mechanics are described by differential
equations on a function space ${\cal M}$:
\[ \frac{d}{dt}A_t(q,p)={\cal L} A_t(q,p) \ , \quad
\frac{d}{dt} \rho_t(q,p)=\Lambda \rho_t(q,p) \ . \]
The operators ${\cal L}$ and  $\Lambda$,
which act on the elements of function space ${\cal M}$, define dynamics.
These operators are infinitesimal generators of dynamical semigroups
and are called dynamical operators.
The first equation describes
evolution of an observable in the Hamilton picture,
and the second equation describes evolution of
a state in the Liouville picture.

Dynamics of an observable $\hat A_t(\hat q,\hat p)$ and a state $\hat \rho_t$
in quantum mechanics
are described by differential equations on an operator space $\hat{\cal M}$:
\[ \frac{d}{dt} \hat A_t( \hat q, \hat p)=
\hat {\cal L} \hat A_t(\hat q, \hat p) \ ,
\quad \frac{d}{dt} \hat \rho_t=\hat \Lambda \hat \rho_t \ . \]
Here $\hat {\cal L}$ and $\hat \Lambda$ are superoperators, i.e.
operators act on the elements of operator space $\hat{\cal M}$.
These superoperators are infinitesimal
generators of quantum dynamical semigroups \cite{Koss1,Koss2,Lind1}.
The first equation describes dynamics in the Heisenberg picture,
and the second - in the Schroedinger picture.

It is easy to see that quantization of the dynamical operators
${\cal L}$ and $\Lambda$ leads to dynamical
superoperators $\hat {\cal L}$ and $\hat \Lambda$. Therefore,
generalization of canonical quantization must
map operators into superoperators.

The usual method of quantization is applied to classical systems,
where the dynamical operators have the forms
${\cal L} A(q,p)=\{ A(q,p), H(q,p)\}$ and
$\Lambda \rho(q,p)=-\{ \rho(q,p), H(q,p)\}$.
Here the function $H(q,p)$ is an observable
which characterizes dynamics and is called the Hamilton function.
Quantization of a dynamical operator which
can be represented as Poisson bracket with a function
is defined by the usual canonical quantization.
Quantization of  real functions $A(q,p)$ and $H(q,p)$ usually
leads to self-adjoint operators $\hat A(\hat q, \hat p)$ and
$\hat H(\hat q, \hat p)$.
Quantization of the Poisson bracket $\{A(q,p),H(q,p)\}$ usually defines as
commutator $(i/\hbar)[\hat H(\hat q, \hat p), \hat A(\hat q, \hat p)]$.
Therefore quantization of these dynamical operators
is uniquely defined by the usual canonical quantization.

Quantization of a dissipative and non-Hamiltonian classical system
by using Hamiltonian meets ambiguities which follow from the problems
of variational description of these systems  \cite{DMS,Hen}.
Quantization of dissipative and non-Hamiltonian systems is not defined by
the usual canonical quantization. Therefore, it is necessary to consider
some generalization of canonical quantization.
A generalized procedure must define a map
of operator into superoperator \cite{prep20,qfthep}.
The usual canonical quantization of observables 
must be derived as a specific case
of generalized quantization
for quantization of operator of multiplication on a function.

In this paper Weyl quantization of dissipative and non-Hamiltonian
classical systems is considered. Generalization of canonical Weyl
quantization, which maps an evolution equation on a function space into
an evolution equation on an operator space, is suggested. 
An analysis of generalized
Weyl quantization is performed for operator, which
cannot be represented as the Poisson bracket with some Hamilton function.

\section{Canonical Weyl Quantization}

Let us consider main points of the usual method of canonical
quantization \cite{Ber1,BJ,kn1,kn2}.
Let $q_k$ be canonical coordinates and $p_k$
are canonical momenta, where $k=1,...,n$.
The basis of the space ${\cal M}$ of
functions $A(q,p)$ is defined by functions
\be
\label{f1}
W(a,b,q,p)=e^{(i/\hbar)(aq+bp)} \ ,
\quad aq=\sum^{n}_{k=1} a_k q_k  \ .
\ee
Quantization transforms coordinates $q_k$ and momenta $p_k$ to
operators $\hat q_k$ and $\hat p_k$. Weyl quantization of the basis
functions (\ref{f1}) leads to the Weyl operators
\be
\label{f2} \hat W(a,b,\hat q, \hat p)=e^{
(i/\hbar)(a\hat q+b \hat p)} \ ,
\quad a \hat q=\sum^{n}_{k=1} a_k \hat q_k \ .
\ee
Operators (\ref{f2}) form a basis of the operator space $\hat {\cal M}$.
Classical observable, characterized by the function $A(q,p)$,
can be represented in the form
\be
\label{f3}
A(q,p)=\frac{1}{(2\pi \hbar)^n} \int \tilde A(a,b) W(a,b,q,p) d^n a d^n b \ ,
\ee
where $\tilde A(a,b)$ is the Fourier image of the function $A(q,p)$.
Quantum observable $\hat A(\hat q,\hat p)$ which corresponds to $A(q,p)$
is defined by formula
\be
\label{f4} \hat A(\hat q, \hat p)=\frac{1}{(2\pi \hbar)^n}
\int \tilde A(a,b) \hat W(a,b,\hat q,\hat p) d^na d^nb  \ .
\ee
This formula can be considered as an operator expansion for
$\hat A(\hat q, \hat p)$ in the operator basis (\ref{f2}).
The direct and inverse Fourier transformations allow
to write the formula (\ref{f4}) for the operator $\hat A(\hat q, \hat p)$ as
\[ \hat A(\hat q,\hat p)=\frac{1}{(2\pi \hbar)^{2n}} \int A(q,p) \]
\be \label{f41}
\times \hat W(a,b, \hat q - q \hat I,\hat p - p \hat I)
d^na d^nb d^nq d^np \ .
\ee
The function $A(q,p)$ is called the Weyl symbol of the
operator $\hat A (\hat q, \hat p)$.
Canonical quantization defined by (\ref{f41}) is called the Weyl
quantization. The Weyl operator (\ref{f2}) in formula (\ref{f5}) leads
to Weyl quantization. Another basis operator leads to different
quantization scheme \cite{BJ}.

The correspondence between operators and symbols completely
is defined by formulas which express symbols of operators
$\hat q_{k} \hat A$,
$\hat A \hat q_{k}$, $\hat p_{k} \hat A$, $\hat A \hat p_{k}$
($ k=1,...,n $) throught operator symbol $\hat A$.
Weyl quantization  $\pi_W$ can be defined by formulas
\be \label{p1}
\pi_{W}((q_{k}+\frac{i \hbar}{2} \frac{\partial}{\partial p_{k}})A(q,p))=
\hat q_{k} \hat A \ , \ee
\be \label{p11}
\pi_{W}((q_{k}-\frac{i \hbar}{2} \frac{\partial}{\partial p_{k}})A(q,p))=
\hat A \hat q_{k} \ , \ee
\be \label{p2}
\pi_{W}((p_{k}-\frac{i \hbar}{2} \frac{\partial}{\partial q_{k}})A(q,p))=
\hat p_{k} \hat A \ , \ee
\be \label{p21}
\pi_{W}((p_{k}+\frac{i \hbar}{2} \frac{\partial}{\partial q_{k}})A(q,p))=
\hat A \hat p_{k} \ , \ee
for all $\hat A=\pi_{W}(A(q,p))$.
Proof of these formulas is contained in \cite{Ber3}.
We obviously have
\be \label{p3}
\pi_{W}(\frac{\partial}{\partial q_{k}}A(q,p))=
-\frac{1}{i \hbar}(\hat p_{k} \hat A-\hat A \hat p_{k}) \ , \ee
\be \label{p31}
\pi_{W}(\frac{\partial}{\partial p_{k}}A(q,p))=
\frac{1}{i \hbar}(\hat q_{k} \hat A-\hat A \hat q_{k}) \ , \ee
\be \label{p4}
\pi_{W}(q_{k}A(q,p))=\frac{1}{2}(\hat q_{k} \hat A+\hat A \hat q_{k}) \ ,
\ee
\be \label{p41}
\pi_{W}(p_{k}A(q,p))=\frac{1}{2}(\hat p_{k} \hat A+\hat A \hat p_{k}) \ .
\ee

Algebraic structures can be defined on the set of obrervables.
Lie algebra, Jordan algebra and $C^{*}$-algebra are usually
considered on the spaces ${\cal M}$ and $\hat {\cal M}$.

Lie algebra $L({\cal M})$ on the set ${\cal M}$ is defined by
Poisson bracket
\[ \{A(q,p),B(q,p)\}= \]
\be \label{f5} =\sum^n_{k=1} \Bigl( \frac{\partial
A(q,p)}{\partial q_k} \frac{\partial B(q,p)}{\partial p_k} -
\frac{\partial A(q,p)}{\partial p_k}
\frac{\partial B(q,p)}{\partial q_k} \Bigr) \ .
\ee
Quantization of the Poisson bracket usually defines
as self-adjoint commutator
\[ \frac{1}{i\hbar} [ \hat A(\hat q, \hat p),
\hat B( \hat q, \hat p)]= \]
\be \label{f6}
=\frac{1}{i\hbar} \Bigl( \hat A(\hat q, \hat p) \hat B(\hat q, \hat p)
- \hat B( \hat q, \hat p) \hat A(\hat q, \hat p) \Bigr) \ .
\ee
The commutator defines Lie algebra $\hat L(\hat {\cal M})$
on the set $\hat {\cal M}$.
Leibnitz rule is satisfied for the Poisson brackets.
As a result, the Poisson brackets are defined by basis Poisson brackets
for canonical coordinates and momenta
\[ \{ q_k , q_m\}=0 , \ \ \{ p_k, p_m \}=0  ,
\ \ \{q_k, p_m\}=\delta_{km} \ . \]
Quantization of these relations lead to the canonical commutation relations
\be
\label{ccr}
[\hat q_k, \hat q_{m}]=0 , \ \ [\hat p_k, \hat p_{m}]=0 ,
\ \ [\hat q_k, \hat p_{m}]=i \hbar \delta_{km} \hat I \ .
\ee
These relations define $(2n+1)$-parametric Lie algebra
$\hat L(\hat {\cal M})$, called Heisenberg algebra.

Jordan algebra $J({\cal M})$ for the set ${\cal M}$ is defined by the
multiplication $A \circ B$ which coincides with the usual associative
multiplication of functions.
Weyl quantization of the Jordan algebra $J({\cal M})$ leads to the operator
special Jordan algebra $\hat J(\hat {\cal M})$ with multiplication
\[ [\hat A, \hat B]_{+}=\hat A \circ \hat B=
\frac{1}{4}[(\hat A+\hat B)^2-(\hat A-\hat B)^2 ] \ . \]
Jordan algebra for classical observables is associative algebra, 
that is, all associators are equal to zero:
\[ (A\circ B) \circ C-A \circ (B \circ C)=0 \ . \]
In general case Jordan algebra associator for quantum observables
is not equal to zero
\be \label{NAR} (\hat A \circ \hat B) \circ \hat C- 
\hat A \circ (\hat B \circ \hat C)=
\frac{1}{4} [\hat B, [\hat C, \hat A]] \ . \ee
This nonassociativity of the operator Jordan algebra  $\hat J(\hat {\cal M})$
leads to the ambiguity of canonical quantization.
The arbitrariness is connected with ordering of noncommutative opetators.

$C^{*}$-algebra can be defined on the set of quantum observables
described by the bounded linear operators.
In general case an operator which is a result of associative multiplication
of the self-adjoint operators is not self-adjoint operator. Therefore, 
quantization of multiplication of classical observables does not lead to
multiplication of the correspondent quantum observables.
Universal enveloping algebra $\hat U(\hat L)$
for the Lie algebra $\hat L(\hat {\cal M})$ which is generated by
commutation relations (\ref{ccr}) usually is considered
as associative algebra \cite{kn1,kn2}.

Let us consider a classical dynamical system
defined by Hamilton function $H(q,p)$.
Usually the quantization procedure is applied to classical systems
with dynamical operator
\[ {\cal L}=-\{H(q,p), \ . \ \}= \]
\be \label{f6'}
=-\sum^n_{k=1} \Bigl( \frac{\partial
H(q,p)}{\partial q_k} \frac{\partial}{\partial p_k} -
\frac{\partial H(q,p)}{\partial p_k}
\frac{\partial}{\partial q_k} \Bigr) \ .
\ee
Here $H(q,p)$ is an observable which defines dynamics of a classical system.
The observable $H(q,p)$ is called the Hamilton function.
The time evolution of a classical observable is described by
\be
\label{eq1}
\frac{d}{dt} A_{t}(q,p)=\{ A_{t}(q,p), H(q,p) \} \ .
\ee
If the dynamical operator has form (\ref{f6'}),
then system is Hamiltonian system.
Weyl quantization of the functions $A_{t}(q,p)$ and $H(q,p)$ leads to
operators  $\hat A_{t}(\hat q, \hat p)$ and $\hat H(\hat q,\hat p)$.
Usually a quantization of
Poisson bracket \ $\{A_{t}(q,p),H(q,p)\}$ defines as 
$(i/\hbar)[ \hat H(\hat q,\hat p), \hat A_{t}(\hat q, \hat p)]$.
Finally canonical quantization of equation (\ref{eq1}) 
leads to the Heisenberg equation
\[ \frac{d}{dt} \hat A_t(\hat q, \hat p) =
\frac{i}{\hbar} [\hat H( \hat q, \hat p),\hat A_{t} (\hat q, \hat p) ] \ . \]
Therefore, canonical quantization of dynamical operator (\ref{f6'}) 
defines as superoperator
\be
\label{f7}
\hat {\cal L}=\frac{i}{\hbar} [\hat H( \hat q, \hat p), \ . \ ]=
\frac{i}{\hbar}(\hat H^l( \hat q, \hat p)-\hat H^r( \hat q, \hat p) )
\ee
Here left and right superoperators $\hat H^l ( \hat q, \hat p)$ and
$\hat H^r(\hat q, \hat p)$ correspond to Hamilton operator
$\hat H(\hat q, \hat p)$.
These superoperators are defined by formulas
\[ \hat H^l \hat A = \hat H \hat A \ , \quad
\hat H^r \hat A= \hat A \hat H \ .  \]
Note that {\it a result of Weyl quantization (\ref{p3}), (\ref{p41}) 
for the Poisson bracket $\{A(q,p),B(q,p)\}$ in general case 
is not a commutator $(-i/\hbar)[\hat A(\hat q, \hat p),
\hat B(\hat q, \hat p)]$.}

Quantization of dynamical operator, which can be represented as Poisson
bracket with a function, is defined by canonical quantization.
Therefore, quantization of Hamiltonian systems can be completely
defined by the usual method of quantization.

\section{General Dynamical System}

Let us consider the time evolution of classical observable $A_t(q,p)$,
described by the general differential equation
\[ \frac{d}{dt} A_t(q,p)={\cal L}(q,p,\partial_q,\partial_p) A_t(q,p) \ , \]
where
\[ \partial_{q}=\frac{\partial}{\partial q} \ , \quad
 \partial_{p}=\frac{\partial}{\partial p} \ . \]
Here ${\cal L}(q,p,\partial_q,\partial_p)$ is an operator
on the function space ${\cal M}$.
In general case this operator cannot be expressed by Poisson bracket
with a function $H(q,p)$.
We would like to generalize the quantization procedure from the dynamical
operators (\ref{f6'}) to general operators  
${\cal L}= {\cal L}(q, p, \partial_q, \partial_p)$.
In order to describe generalized quantization we must define
a general operator ${\cal L}(q,p,\partial_q,\partial_p)$
using some operator basis.
For simplicity, we assume that operator
${\cal L}(q,p,\partial_q,\partial_p)$ is a bounded operator.

Let us define the basis operators which generate the dynamical
operator ${\cal L}={\cal L}(q,p,\partial_q,\partial_p)$.
Operators $Q^k_1$ and $Q^k_2$ are operators of multiplication
on $q_k$ and $p_k$.
Operators $P^k_1$ and $P^k_2$ are self-adjoint differential operators
with respect to $q_k$ and $p_k$, that is $P^k_1=-i\partial / \partial q_k$
and  $P^k_2=-i \partial/ \partial p_k$.
These operators obey the conditions:\\
1. $Q^k_1 1=q_k$, \ $Q^k_2 1=p_k$ \ and  \ $P^k_1 1=0$, \ $P^k_2 1=0$.\\
2. $(Q^k_{1,2})^*=Q^k_{1,2}$, \ $(P^k_{1,2})^*=P^k_{1,2}$. \\
3. $P^{k}_{1,2}(A\circ B)=(P^{k}_{1,2}A)\circ B+A\circ (P^{k}_{1,2}B)$.\\
4. $[Q^k_{1,2}, P^m_{1,2}]=i\delta_{km}$, \ $[Q^k_{1,2},P^m_{2,1}]=0$. \\
5. $[Q^k_{1,2},Q^m_{1,2}]=0$, \ $[P^k_{1,2},P^m_{1,2}]=0$ .\\
Conjugation operation is defined with respect to the usual scalar product
of function space.
Commutation relations for the operators $P^{k}_{1,2}$ and $Q^{k}_{1,2}$
define $(4n+1)$-parametric Lie algebra. These relations are analogous
to canonical commutation relations (\ref{ccr}) for
$\hat q_{k}$ and $\hat p_{k}$ with double numbers of degrees of freedom.

Operators $Q^k_{1,2}$ and $P^k_{1,2}$ allow to introduce operator basis
\[ V(a_1,a_2,b_1,b_2,Q_1,Q_2,P_1,P_2)= \]
\be \label{V1}
=exp\{ i(a_1Q_1+a_2Q_2+b_1P_1+b_2 P_2)\} \ ,
\ee
for the space ${\cal A}({\cal M})$ of dynamical operators.
These basis operators are analogous to the  Weyl operators (\ref{f2}).
Note that basis functions (\ref{f1}) can be derived from 
operators (\ref{V1}) by the formula
\[ W(a,b,q,p)= \]
\[ =V((a/\hbar),(b/\hbar),0,0,Q_1,Q_2,P_1,P_2)1 \ . \]

The algebra ${\cal A}({\cal M})$ of bounded dynamical operators
can be defined as $C^{*}$-algebra, generated
by $Q^{k}_{1,2}$ and $P^{k}_{1,2}$.
It contains all operators (\ref{V1}) and is closed for linear
combinations of (\ref{V1}) in operator norm topology.
A dynamical operator ${\cal L}$ can be defined as an operator function
of basis operators $Q^k_{1,2}$ and $P^k_{1,2}$:
\[ {\cal L}(Q_1,Q_2,P_1,P_2)=\frac{1}{(2\pi)^{2n}} \int
\tilde L(a_1,a_2,b_1,b_2) \]
\be \label{f9}
\times e^{i(a_1Q_1+a_2Q_2+b_1P_1+b_2P_2)} d^na_1 d^na_2 d^nb_1 d^n b_2 \ ,
\ee
where $\tilde L(a_1,a_2,b_1,b_2)$ is square-integrable
function of real variables $a_{1,2}$ and $b_{1,2}$.
The function $\tilde L(a_1,a_2,b_1,b_2)$ is Fourier image of the
symbol of operator ${\cal L}(q,p,\partial_q,\partial_p)$.
The set of bounded operators ${\cal L}(Q_1,Q_2,P_1,P_2)$
and their uniformly limits
forms the algebra ${\cal A}({\cal M})$ of dynamical operators.

\section{Weyl Quantization of \\ Basis Operators}

To define a quantization of dynamical operator ${\cal L}$ we need
to describe quantization of the operators $Q^k$ and $P^k$.
Let us require that the superoperators $\hat Q^k$
and $\hat P^k$ satisfy the relations which are the
quantum analogs to the relations for
the operators $Q^k$ and $P^k$:\\
1. $\hat Q^k_1 \hat I=\hat q_k$, \ $\hat Q^k_2 \hat I=\hat p_k$, and
$\hat P^k_{1,2} \hat I=0$.\\
2. $(\hat Q^k_{1,2})^*=\hat Q^k_{1,2}$, \
$(\hat P^k_{1,2})^*=\hat P^k_{1,2}$. \\
3. $\hat P^{k}_{1,2}(\hat A \circ \hat B)=
(\hat P^{k}_{1,2} \hat A) \circ \hat B+
\hat A \circ (\hat P^{k}_{1,2} \hat B)$.\\
4. $[\hat Q^k_{1,2},\hat P^m_{1,2}]=i\delta_{km} \hat I$,
\ $[\hat Q^k_{1,2},\hat P^m_{2,1}]=0$. \\
5. $[\hat Q^k_{1,2},\hat Q^m_{1,2}]=0$, \
$[\hat P^k_{1,2},\hat P^m_{1,2}]=0$.\\
Superoperator $\hat {\cal L}$ is called self-adjoint, if the relation
$<\hat {\cal L}\hat A|\hat B>=<\hat A|\hat {\cal L} \hat B>$ is satisfied.
The scalar product $<\hat A|\hat B>$ on the operator space ${\cal M}$
is defined by the relation $<\hat A|\hat B> \equiv Sp[\hat A^* \hat B]$.
An operator space with this scalar product is called Liouville space
\cite{kn1,kn2}.

To quantize the operator $P^k_{1,2}$ we use the relations
\[ P^k_1 A(q,p)=-i\frac{\partial}{\partial q_k} A(q,p)
=i\{p_k,A(q,p)\} \ , \]
\[ P^k_2 A(q,p)=-i\frac{\partial}{\partial p_k} A(q,p)
=-i\{q_k,A(q,p)\} \ . \]
Weyl quantization (\ref{p3},\ref{p31}) of these expressions lead to
\[ \hat P^k_1 \hat A(\hat q, \hat p)=
\frac{1}{ \hbar}[\hat p_k, \hat A(\hat q, \hat p) ] \ , \]
\[ \hat P^k_2 \hat A(\hat q, \hat p)=-
\frac{1}{ \hbar}[\hat q_k, \hat A(\hat q, \hat p) ] \ . \]
As a result, we obtain
\be
\label{Pk1}
\hat P^k_1 =\frac{1}{ \hbar}[\hat p_k, \ . \ ]
= \frac{1}{\hbar}(\hat p_k^l-\hat p_k^r) \ .
\ee
\be
\hat P^k_2 =- \frac{1}{ \hbar}[\hat q_k, \ . \ ]
=- \frac{1}{\hbar}(\hat q_k^l-\hat q_k^r) \ ,
\ee

Here we use superoperators
$\hat q_k^l$,  $\hat q_k^r$ and $\hat p_k^l$,  $\hat p_k^r$
which satisfy the non-zero commutation relations
\[ [\hat q_k^l,\hat p_m^l]=i \hbar \delta_{km} \hat I , \ \quad
[\hat q_k^r,\hat p_m^r]=-i \hbar \delta_{km} \hat I  . \  \]
These relations follow from canonical commutation relations (\ref{ccr}).

To quantize the operator $Q^k_{1,2}$ we use formulas (\ref{p4}), (\ref{p41}).
It is known \cite{Ber3,Ber2} that Weyl quantization (\ref{p4}), (\ref{p41})
of the expressions $q_k \circ A(q,p)$ and $p_k \circ A(q,p)$
leads to $\hat q_k \circ \hat A(\hat q, \hat p)$
and $\hat p_k \circ \hat A(\hat q, \hat p)$.
Therefore, Weyl quantization of the operators $Q^k_{1,2}$ lead
to superoperators
\be \label{Qk1}
\hat Q^k_1=[\hat q_{k}, \ . \ ]_{+}=
\frac{1}{2} (\hat q_k^l+\hat q_k^r) \ ,
\ee
\be \label{Qk2}
\hat Q^k_2=[\hat p_{k}, \ . \ ]_{+}=
\frac{1}{2} (\hat p_k^l+\hat p_k^r) \ ,
\ee
where  $\hat Q^k_1 \hat A=\hat q_k \circ \hat A$ and
$\hat Q^k_2 \hat A=\hat p_k \circ \hat A$.

{\it If the Weyl quantization for observables is considered then we 
must consider the Weyl quantization for dynamical operators.}
The Weyl quantization leads only to this form (\ref{Qk1}), (\ref{Qk2})
of superoperators $\hat Q^k_{1,2}$. The other quantization
of the observables \cite{BJ,Ber2} leads to other form of
the superoperators $\hat Q^k_{1,2}$.


Weyl quantization of the basis operators (\ref{V1}) leads to the basis superoperators
\[ \hat V(a_1,a_2,b_1,b_2,\hat Q_1,\hat Q_2, \hat P_1, \hat P_2)= \]
\be \label{HV1}
=exp\{ i(a_1 \hat Q+a_2\hat Q_2+b_1 \hat P_1+b_2 \hat P_2)\} \ .
\ee

\section{Weyl Quantization of \\ Operator Function}

Let us consider the dynamical operator ${\cal L}$
as a function of the basis operators $Q^k_{1,2}$ and $P^k_{1,2}$.
Generalized Weyl quantization can defined as a map from
dynamical operator space $A({\cal M})$ to dynamical superoperator
space $\hat A(\hat {\cal M})$.
This quantization of the operator
\[ {\cal L}(Q_1,Q_2,P_1,P_2)=\frac{1}{(2\pi)^{2n}} \int \tilde
L(a_1,a_2,b_1,b_2) \]
\[ \times e^{i(a_1Q_1+a_2Q_2+b_1P_1+b_2P_2)} d^na_1 d^na_2 d^nb_1 d^nb_2 \ , \]
leads to the corresponding superoperator
\[ \hat {\cal L}(\hat Q_1,\hat Q_2,\hat P_1, \hat P_2)=
\frac{1}{(2\pi)^{2n}} \int \tilde L(a_1,a_2,b_1,b_2) \]
\be \label{f11}
\times e^{i(a_1 \hat Q_1+a_2 \hat Q_2+b_1 \hat P_1+ b_2 \hat P_2)} d^na_1
d^na_2 d^nb_1 d^nb_2 \  .
\ee

If the function $\tilde L(a_1,a_2,b_1,b_2)$ is connected with Fourier image
$\tilde A(a_1,a_2)$ of the function $A(q,p)$ by the relation
\[ \tilde L(a_1,a_2,b_1,b_2)=
(2 \pi)^n \delta (b_1) \delta (b_2) \tilde A(a_1,a_2) \ , \]
then formula (\ref{f11}) defines the Weyl quantization
of the function $A(q,p)={\cal L}(Q_1,Q_2,P_1,P_2)1$ by the relation
\[ \hat A(\hat q,\hat p)=
\hat {\cal L}(\hat Q_1,\hat Q_2,\hat P_1, \hat P_2) \hat I \ . \]
Here we use $\hat Q^k_1 \hat I=\hat q_k$ and $\hat Q^k_2 \hat I=\hat p_k$.
Therefore {\it the usual Weyl quantization is a spesific case
of suggested quantization procedure.}


Superoperators $\hat Q^k_{1,2}$ and $\hat P^k_{1,2}$
can be represented by
$\hat q_k^l$, $\hat q_k^r$ and $\hat p_k^l$, $\hat p_k^r$.
Formula (\ref{f11}) is written in the form
\[ \hat {\cal L}( \hat q^l, \hat q^r,\hat p^l, \hat p^r)=
\frac{1}{(2\pi)^{2n}} \int L({a}_1,{a}_2,{b}_1,{b}_2)  \]
\[ \times W^l({a}_1,{a}_2, \hat q, \hat p)W^r({b}_1,{b}_2, \hat q, \hat p)
d^{n}{a}_1 d^{n}{a}_2 d^{n}{b}_1 d^{n}{b}_2 \ . \]
Here $W^l(a,b, \hat q, \hat p)$ and $W^l(a,b, \hat q,\hat p)$ are left and
right superoperators corresponding to the Weyl operator (\ref{f2}).
These superoperators can be defined by
\[ W^{l}(a,b,\hat q,\hat p)=W(a,b,\hat q^l,\hat p^l) \ , \]
\[ W^{r}(a,b,\hat q,\hat p)=W(a,b,\hat q^r,\hat p^r) \ . \]


We can derive \cite{prep20} a relation which represents the superoperator
$\hat{\cal L}$ by operator ${\cal L}$.
Let us write the analog of relation (\ref{f41})
between an operator $\hat A$ and a function $A$.
To simplify formulas, we introduce new notations.
Let $X^s$, where $s=1,...,4n$, denote the operators
$Q^k_{1,2}$ and $P^k_{1,2}$, where $k=1,...,n$, that is
\[ X^{2k-1}=q_k \ , \quad X^{2k}=p_k \ , \]
\[ X^{2k-1+2n}=-i\frac{\partial}{\partial q_k} \ , \quad
X^{2k+2n}=-i \frac{\partial}{\partial p_k} \ . \]
Let us denote the parameters $a^k_{1,2}$ and $b^k_{1,2}$,
where $k=1,...,n$, by $z^s$, where $s=1,...,4n$.
Then formula (\ref{f9}) can be rewritten by
\[ {\cal L}=\frac{1}{(2\pi)^{2n}} \int L(z) e^{izX} d^{4n}z \ . \]
Formula (\ref{f11}) for the superoperator $\hat {\cal L}$
is written in the form
\[ \hat {\cal L}=\frac{1}{(2\pi)^{2n}}
\int L(z) e^{iz \hat X} d^{4n}z \ . \]
The result relation  \cite{prep20} which represents the superoperator
$\hat{\cal L}$ by operator ${\cal L}$ can be written in the form
\[ \hat {\cal L}=\frac{1}{(2\pi)^{4n}}
\int e^{-i\alpha(z+z')} e^{iz\hat X} \]
\be \label{ff3}
Sp[{\cal L} e^{iz' X}] d^{4n}z d^{4n} \alpha d^{4n}z' \ .
\ee

\section{Oscillator with friction}

Let us consider $n$-dimensional oscillator with friction
$ F^{k}_{fric}=-\alpha_{km}p_{m}-\beta_{kms} p_m p_s$.
The time evolution equation for this oscillator has the form
\[ \frac{d}{dt}q_k=\frac{1}{m}p_k \ , \]
\be \label{ho}
\frac{d}{dt} p_k=-(m\omega^2 q_k+\alpha_{km}p_{m}+\beta_{kms} p_m p_s) \ ,
\ee
where $k,m,s=1,...,n$. If $n=3$, $\omega=0$ and non-zero coefficients are
\[ \alpha_{11}=10, \ \alpha_{12}=-10, \ \alpha_{21}=-28, \
\alpha_{22}=1, \ \alpha_{33}=8/3 \ , \]
\[ \beta_{213}=\beta_{231}=0.5, \ \beta_{312}=\beta_{321}=-0.5
\ , \]
then we have the Lorenz system \cite{Lor}
with respect to  $x=p_{1}$, $y=p_{2}$ and $z=p_{3}$.
If non-zero coefficients are
\[ \alpha_{12}=\alpha_{13}=1, \ \alpha_{21}=-1 \ , \]
\[ \alpha_{22}=\alpha_{31}=-0.2, \ \alpha_{33}=5.7 \ , \]
\[ \beta_{313}=\beta_{331}=-0.5 \ , \]
then we obtain the Rossler sytem \cite{Ros}.
For the case
\[ \alpha_{11}=0.4, \ \alpha_{12}=-1, \ \alpha_{21}=1, \ \alpha_{22}=0.4, \]
\[ \alpha_{33}=-\alpha=-0.175, \ \beta_{123}=\beta_{132}=-5 \ , \]
\[ \beta_{213}=\beta_{231}=-2.5, \
\beta_{312}=\beta_{321}=2.5 \ , \]
we have the Leipnik-Newton system \cite{LN}.

The dynamical equation for the classical observable $A_{t}(q,p)$ is written
\[ \frac{d}{dt}A_t(q,p)={\cal L}(q,p,\partial_q, \partial_p ) A_t(q,p) \ . \]
Differentiation of the function $A_{t}(q,p)$ and equations (\ref{ho})
give
\[
\frac{d A_t (q,p)}{dt}=\frac{1}{m} p_k\frac{\partial A_t (q,p)}{\partial q_k}
-m\omega^2 q_k\frac{\partial A_t(q,p)}{\partial p_k}-   \]
\be \label{ho2}
-(\alpha_{km} p_m+\beta_{kms}p_mp_s )\frac{\partial A_t
(q,p)}{\partial p_k}  \ .
\ee
Dynamical operator ${\cal L}(q,p,\partial_q, \partial_p)$
for system (\ref{ho}) has the form
\[ {\cal L}(q,p,\partial_q,\partial_p)=
\frac{1}{m} p_k \frac{\partial}{\partial q_k}-
m\omega^2 q_k \frac{\partial}{\partial p_k}- \]
\be \label{do}
-(\alpha_{km} p_m+\beta_{kms}p_m p_s )\frac{\partial}{\partial p_k} \ .
\ee
This operator can be rewritten in the form
\[ {\cal L}(Q_1,Q_2,P_1,P_2)=
\frac{i}{m} Q^k_2 P^k_1-i m\omega^2 Q^k_1P^k_2- \]
\be \label{LQP}
-i(\alpha_{km} Q^m_2+\beta_{kms}Q^m_2 Q^s_2 )P^k_2 \ .
\ee
If we consider the Weyl quantization for observables
$A(q,p)$ then we must consider the Weyl quantization for
dynamical operators ${\cal L}(q,p,\partial_{q},\partial_{p})$.
The Weyl quantization of operator (\ref{LQP}) leads to superoperator
\[ \hat {\cal L}(\hat Q_1,\hat Q_2,\hat P_1,\hat P_2)=
\frac{i}{m} \hat Q^k_2 \hat P^k_1-i m\omega^2 \hat Q^k_1 \hat P^k_2 - \]
\[ - i(\alpha_{km} \hat Q^m_2+
\beta_{kms}\hat Q^m_2 \hat Q^s_2 )\hat P^k_2 \ . \]
Let us use definitions (\ref{Pk1}), (\ref{Qk1}) of the
operators $\hat P_{1,2}$ and $\hat Q_{1,2}$.
The time evolution equation for a quantum observable $\hat A$
takes the form
\[ \frac{d}{dt}\hat A_{t}=\frac{i}{\hbar}[\hat H, \hat A_t]
+\frac{i}{\hbar}a_{km} \hat p_m \circ [\hat q_k,\hat A_t]+ \]
\be \label{LRL}
+\frac{i}{\hbar} b_{kms}
\hat p_m \circ( \hat p_s \circ [\hat q_k,\hat A_{t}]) \ .
\ee
Here $\hat A \circ \hat B=(1/2)(\hat A\hat B+\hat B\hat A)$ and
\[ \hat H=\frac{\hat p^2}{2m}+\frac{m \omega^2 \hat q^2}{2} \ . \]
Equation (\ref{LRL}) describes  \cite{qfthep,prep21} quantum analogous
of the generalized Lorenz-Rossler-Leipnik-Newton equation (\ref{ho2}).

Note that Weyl quantization of $p_mp_s \{p_k,A_t (q,p)\}$
does not lead to the term $(-i/\hbar)(\hat p_m \circ \hat p_s) \circ [\hat
p_k,\hat A_t]$. It gives the term
$\hat p_m \circ( \hat p_s \circ [\hat q_k,\hat A_{t}])$
and in general case
\[ \hat p_m \circ( \hat p_s \circ [\hat q_k,\hat A])-
(\hat p_m \circ \hat p_s) \circ [\hat q_k,\hat A]=
\frac{1}{4}[\hat p_s, [ \hat p_m,[\hat q_k,\hat A]]] \ . \]

\section{Fokker-Planck-Type System}

Let us consider Liouville operator $\Lambda$, which acts on the
normed distribution density function $\rho(q,p,t)$ and has the form
of second order differential operator
\[ \Lambda=d_{qq} \frac{\partial^{2}}{\partial q^{2}}+
2 d_{qp} \frac{\partial^{2}}{\partial q \partial p}+
d_{pp} \frac{\partial^{2}}{\partial p^{2}}+ \]
\be \label{liu-li} +c_{qq} q \frac{\partial}{\partial q}+
c_{qp} q \frac{\partial}{\partial p}+
c_{pq} p \frac{\partial}{\partial q}+
c_{pp} p \frac{\partial}{\partial p}+h \ .
\ee
Liouville equation
\[ \frac{d \rho(q,p,t)}{d t}= \Lambda \rho(q,p,t) \]
with operator (\ref{liu-li}) is Fokker-Planck-type equation.
Weyl quantization of the Liouville operator (\ref{liu-li}) leads
to completely dissipative superoperator
$\hat \Lambda$, which acts on the matrix density operator
\[ \hat \Lambda=-\frac{i}{\hbar}(\hat H^l-\hat H^r)+ \]
\[ +\frac{1}{2\hbar}
\sum_{j=1,2}\Bigl( (\hat V^l_j-\hat V^r_j)\hat V^{*r}_j-
(\hat V^{*l}_j-\hat V^{*r}_{j}) \hat V^{*l}_j \Bigr) \ , \]
As the result we have the Markovian master equation \cite{Lind1,kn1,Sand}:
\be
\label{8.12}
\frac{d\hat \rho_{t}}{dt}=-\frac{i}{\hbar}[\hat H,\hat \rho_{t}]+
\frac{1}{2\hbar} \sum_{j=1,2}
([ \hat V_{j} \hat \rho_{t}, \hat V_{j}^{*} ]+
[ \hat V_{j}, \hat \rho_{t} \hat V_{j}^{*} ]) \ .
\ee
Here $\hat H$ is Hamilton operator, which has the form
\[ \hat H=\hat H_{1}+\hat H_{2} \ , \quad
\hat H_{1}=\frac{1}{2m}\hat p^2+\frac{m\omega^{2}}{2} \hat q^{2} \ , \]
\[ \hat H_{2}=\frac{\mu}{2}(\hat p \hat q+\hat q \hat p) \ , \]
where
\[ m=-\frac{1}{c_{pq}} \ , \quad \omega^{2}=-c_{qp}c_{pq} \ , \]
\[ \lambda=\frac{1}{2}(c_{pp}+c_{qq}) \ , \quad
\mu=\frac{1}{2}(c_{pp}-c_{qq}) \ . \]
Operators $\hat V_{j}$ in (\ref{8.12}) can be written in the form
$\hat V_{k}=a_{j}\hat p+b_{j}\hat q$, where $j=1,2$, and complex numbers
$a_{j}$, $b_{j}$ satisfy the relations
\[ d_{qq}=\frac{\hbar}{2}\sum_{j=1,2}{\vert a_{j}\vert}^2 \ , \quad
d_{pp}=\frac{\hbar}{2}\sum_{j=1,2}{\vert b_{j}\vert}^2 \ , \]
\[ d_{qp}=-\frac{\hbar}{2}  Re(\sum_{j=1,2}a_{j}^{*}b_{j}) \ ,
\quad \lambda=-Im(\sum_{j=1,2}a_{j}^{*}b_{j}) \ . \]
If $h=-2(c_{pp}+c_{qq})$, then quantum Markovian
equation (\ref{8.12}) becomes \cite{Sand}:
\[ \frac{d\hat \rho_t}{dt}=-\frac{i}{\hbar}[\hat H_1, \hat \rho_t]+ \]
\[ +\frac{i(\lambda-\mu)}{\hbar}[\hat p,\hat q \circ \hat \rho_t]-
\frac{i(\lambda+\mu)}{\hbar}[\hat q,\hat p \circ \hat \rho_t]-  \]
\[ -\frac{d_{pp}}{{\hbar}^2}[\hat q,[\hat q,\hat \rho_t]]-
\frac{d_{qq}}{{\hbar}^2}[\hat p,[\hat p, \hat \rho_t]]
+\frac{2d_{pq}}{{\hbar}^2}[\hat p,[ \hat q,\hat \rho_t]] \ . \]
Here $d_{pp}$, $d_{qq}$, $d_{pq}$ are quantum diffusion coefficients and
$\lambda$ is a friction constant.

\section{Lorenz-Type System}

Let us consider the evolution of a classical observable
$A_{t}(q,p)$ for the Lorenz-type system \cite{qfthep,prep21}:
\[ \frac{d}{dt}A_t(q,p)=-\sigma (q_1- p_1)
\frac{\partial A_{t}(q,p)}{\partial q_1}+
\sigma p_2\frac{\partial A_{t}(q,p) }{\partial q_2}+ \]
\be \label{ls}
+(rq_1-p_1-q_1p_2)\frac{\partial A_{t}(q,p) }{\partial p_1}
-(bp_2-q_1p_1)\frac{\partial A_{t}(q,p) }{\partial p_2} \ . \ee

This equation for observables $x=q_1$, $y=p_1$ and $z=p_2$,
describes the classical Lorenz model \cite{Lor,Spar}:
\[ \frac{dx_t}{dt}=-\sigma x_t+ \sigma y_t \ , \]
\[ \frac{dy_t}{dt}=rx_t-y_t-x_tz_t \ , \]
\[ \frac{dz_{t}}{dt}=-bz_t+x_ty_t \ . \]
The Lorenz model \cite{Lor} is one of the most famous classical
dissipative systems.
This system is described by nonlinear differential equations without
stochastic terms, but the system demonstrates chaotic behaviour and
has strange attractor for $\sigma=10, \ r=28, \ b=8/3$ (see \cite{Lor,Spar}).

The Weyl dynamical quantization of the Lozenz-type equation leads to
the quantum Lorenz-type equation
\[ \frac{d}{dt}\hat A_t=\frac{i}{\hbar}
[\frac{\sigma(\hat p^2_1+ \hat p^2_2)}{2}-\frac{r \hat q^2_1}{2},\hat A_{t}]-
\frac{i\sigma}{\hbar} \hat q_1 \circ [ \hat p_1,\hat A_{t}]+ \]
\[ +\frac{i}{\hbar} \hat p_1 \circ [\hat q_{1}, \hat A_{t}]
+\frac{i}{\hbar}b\hat p_2 \circ [\hat q_2,\hat A_{t}]+ \]
\[ +\frac{i}{\hbar}\hat q_1 \circ (\hat p_2 \circ [\hat q_1, \hat A_{t}])
-\frac{i}{\hbar} \hat q_1 \circ (\hat p_1 \circ [\hat q_2,\hat A_{t}]) \ . \]
Note that Weyl quantization of the term
$q_kp_l\{A(q,p),q_m\}$
leads to the term
$(i/\hbar)\hat q_k \circ (\hat p_l \circ [\hat q_m, \hat A])$,
which is equal to
$(i/\hbar)\hat p_l \circ (\hat q_k \circ [\hat q_m, \hat A])$.
Using relation (\ref{NAR}), we can see that
these terms are not equal to
$(i/\hbar)(\hat q_k \circ \hat p_l) \circ [\hat q_m, \hat A]$.


\section{Conclusions}

Quantization of a dynamical operator which is
represented by Poisson bracket with the Hamilton function,
can be defined by the usual canonical quantization. Quantization of a general
dynamical operator for non-Hamiltonian system cannot be described by
usual canonical quantization procedure. We suggest the quantization scheme
which allows to derive quantum analog for the classical
non-Hamiltonian systems.
Relations (\ref{f11}) and (\ref{ff3}) map the operator
${\cal L}(q,p,\partial_{q},\partial_{p})$ which
acts on the functions $A(q,p)$ to the superoperator $\hat {\cal L}$,
which acts on the elements $\hat A(\hat q,\hat p)$ of operator space.
If the operator ${\cal L}$ is an operator of
multiplication on the function $A(q,p)={\cal L}1$, then
formula (\ref{ff3}) defines the usual Weyl quantization of the function
$A(q,p)$ by the relation $\hat A=\hat {\cal L} \hat I$.
Therefore, the usual Weyl quantization of observables is a specific case 
of suggested generalization of Weyl quantization.
The suggested approach allows to derive quantum
analogs of chaotic dissipative systems with strange attractors
\cite{qfthep,prep21}. \\

This work was partially supported by the RFBR grant No. 00-02-17679.



\end{document}